# Bragg Coherent Modulation Imaging: Strain- and Defect-Sensitive Single Views of Extended Samples

A. Ulvestad[1]*, W. Cha[2], I. Calvo-Almazan[1], S. Maddali[1], S. M. Wild[3], E. Maxey[2], M. Dupraz[4], and S. O. Hruszkewycz[1]

[1]Materials Science Division, Argonne National Laboratory, Argonne, Illinois 60439, USA

[2]Advanced Photon Source, Argonne National Laboratory, Argonne, Illinois 60439, USA

[3]Mathematics and Computer Science Division, Argonne National Laboratory, Argonne, Illinois 60439, USA

[4]Paul Scherrer Institute, Switzerland

*aulvestad@anl.gov

**Abstract**

Nanoscale heterogeneity (including size, shape, strain, and defects) significantly impacts material properties and how they function. Bragg coherent x-ray imaging methods have emerged as a powerful tool to investigate, in three-dimensional detail, the local material response to external stimuli in reactive environments, thereby enabling explorations of the structure-defect-function relationship at the nanoscale. Although progress has been made in understanding this relationship, coherent imaging of extended samples is relatively slow (typically requiring many minutes) due to the experimental constraints required to solve the phase problem. Here, we develop Bragg coherent modulation imaging (BCMI), which uses a modulator to solve the phase problem thereby enabling fast, local imaging of an extended sample. Because a known modulator is essential to the technique, we first demonstrate experimentally that an unknown modulator structure can be recovered by using the exit wave diversity that exists in a standard Bragg coherent diffraction imaging (BCDI) experiment. We then show with simulations that a known modulator constraint is sufficient to solve the phase problem and enable a single view of an extended sample that is sensitive to defects and dislocations. Our results pave the way for BCMI investigations of strain and defect dynamics in complex extended crystals with temporal resolution limited by the exposure time of a single diffraction pattern.

**Introduction**

Coherent diffraction imaging (CDI) with x-rays has found diverse applications in fields from biology to materials science[1–4]. In the Bragg geometry, the technique can image the three-dimensional (3D) strain field, which yields insight into the impact of strain on material performance, in isolated (smaller than the x-ray beam)[5–7] and extended (larger than the x-ray beam) samples[8–10]. Bragg CDI experiments are also sensitive to dislocations, which play an important role in polycrystalline materials[11], ion intercalation[12], and phase transformation dynamics[13–15]. As such, this class of experimental methods are important tools in understanding how strain and defects influence and govern material properties in reactive environments[16–19]. All of these methods rely on iterative phase retrieval to solve the phase problem and form real space images[20–22]. For extended crystals, ptychography[23,24] constrains the phase problem by measuring diffraction patterns from overlapping scan positions[20,25–27]. Unfortunately, this approach has limited temporal temporal resolution because the sample dynamics must be constant during the diffraction pattern acquisition at the overlapping positions.

In order to overcome this limitation, recent efforts have focused on using compressed sensing, sparsity, and wavefront modulation to image targeted areas of extended objects from a single diffraction pattern[28–32]. In particular, recent work demonstrated extended sample imaging in a transmission geometry by using a modulator-based approach where an object (the modulator) is placed in between the sample and the detector[30]. Knowledge of the object structure serves as a constraint during iterative phase retrieval: consequently, diffraction patterns from overlapping scan positions are not required. However, the transmission geometry is not sensitive to strain or dislocations. Here we develop a modulator-based approach for the Bragg geometry, including determination of the modulator structure, such that fast, local strain and dislocation sensitive images can be obtained for extended samples. We call this Bragg coherent modulation imaging (BCMI).

In BCMI, the temporal resolution is limited by the acquisition time of a single diffraction pattern, typically on the order of 0.5s. Because the modulator structure, represented as a complex image with amplitude and phase, must be known, we first demonstrate experimentally that an unknown modulator can be recovered by using an isolated crystal as a probe. We then show through simulation that the modulator knowledge is able to solve the extended sample phase problem.

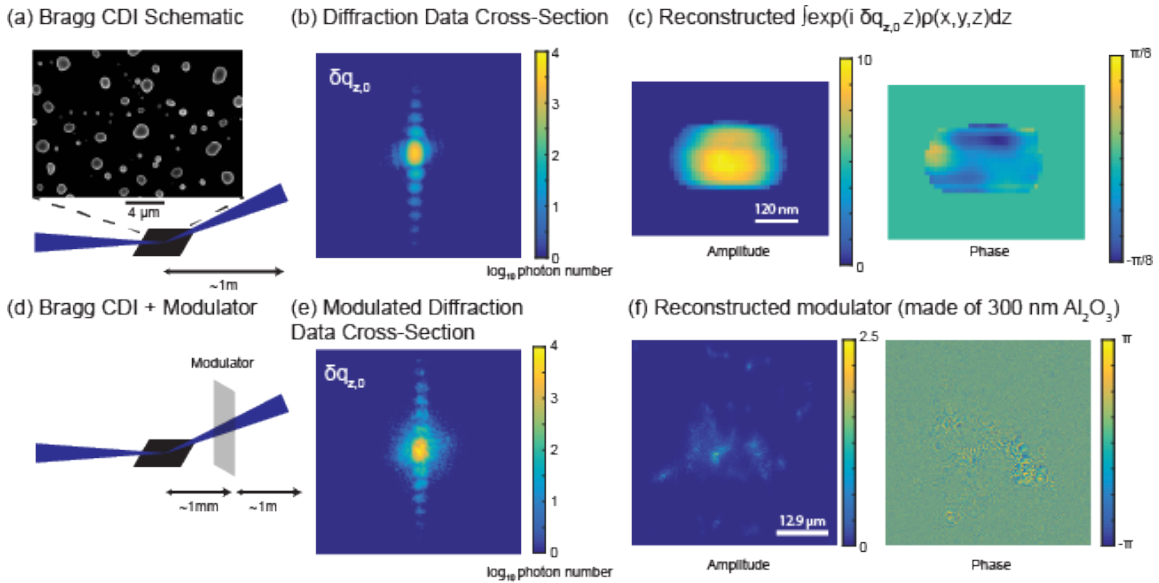

**Figure 1.** Recovering an unknown modulator using a known crystal. **(a)** A monochromatic, spatially coherent x-ray beam is incident on the sample, which is composed of isolated gold nanocrystals on a Si substrate. The beam is 2-4x larger than the nanocrystals. **(b)** 2D cross-section of the 3D experimental diffraction data. The 3D data is collected by rotating the nanocrystal a total of ~0.6° with respect to the incident x-

ray beam. The 2D cross-section shown is the cross-section with the maximum intensity, defined to be at $\delta q_{z,0}$. 121 cross-sections were measured. **(c)** The complex 2D crystal projection, computed from the 3D reconstruction of the data in **(b)**. The 2D image is an integral in the z (exit beam) direction with the appropriate phase term and can be computed after 3D iterative phase retrieval. **(d)** The schematic for determining the modulator. A modulator, consisting of dispersed 0.3 micron $Al_2O_3$ micropolish, has been inserted in between the sample and the detector. **(e)** The 2D experimental diffraction data at $\delta q_{z,0}$ with the modulator inserted. **(f)** The reconstructed complex image of the modulator. The reconstruction algorithm relies on the exit wave diversity to constrain the modulator image. The modulator was reconstructed by using the new algorithm discussed in the text and the known nanocrystal reconstruction from a standard BCDI experiment.

BCMI requires a known modulator to solve the phase problem. While the modulator could be fabricated to have a known structure, which we discuss later, we also demonstrate how an unknown modulator structure can be reconstructed using a known crystal (Fig. 1). Borrowing from ideas in ptychography, we use the diversity in the known crystal's exit wave to constrain the problem (Supplementary Fig. 1). The known crystal's exit wave is determined using a standard BCDI experiment (Fig. 1a). A focused, monochromatic, spatially coherent x-ray beam was used to illuminate a gold nanocrystal and was aligned such that a {111} Bragg peak illuminated the area detector. 3D diffraction data (composed of 121 2D cross-sections or slices) was collected from the nanocrystal as a function of angle over the rocking curve and the central cross-section is shown in Fig. 1b. We cast each angle, *i*, of the rocking curve in terms of deviation from the Bragg condition in the reciprocal space coordinate $\delta q_{z,i}$ that is normal to the detector plane, with $\delta q_{z,0}$ corresponding to the Bragg peak maximum. Iterative phase retrieval was used to recover the 3D complex image of the crystal from these unmodulated data using standard methods (see Methods). During standard iterative phase retrieval, 3D Fourier transforms are used. With the insertion of the modulator, however, every 2D diffraction pattern interacts with the modulator in the near field. Modeling this interaction requires Fresnel propagation of the exit wave at each rocking angle to the modulator. As such, each diffraction cross-section must be considered independently and related to the appropriate 2D crystal projection (Fig. 1c). At the modulator plane, which has a Fresnel number of approximately unity in this experiment and is thus in the near field, the relationship between a particular 2D cross-section defined at a given $\delta q_{z,i}$, and the complex 3D crystal $\rho(x,y,z)$ is given by[33–35]

$$U_i(q_x, q_y, \delta q_{z,i}) = \mathcal{F}r_{2D}( \int P(x,y,z) \cdot e^{i\delta q_{z,i} z} \rho(x,y,z) dz), \quad (1)$$

where $\delta q_{z,i}$ corresponds to the deviation from the Bragg peak maximum, and $P$ is the probe. The probe is set to unity for the modulator reconstruction, but will be required for the extended sample discussion later. $\mathcal{F}r_{2D}$ is the Fresnel propagator. Thus, we can consider the 3D object to be represented as a set of appropriate 2D projections. When a modulator is placed between the sample and the detector (Fig. 1d), these 2D exit waves (Supplementary Fig. 1) interact with the modulator and then propagate (modeled using the Fourier transform) to the detector. The modulator introduces spatial phase and amplitude diversity into the diffracting wavefield and consequently the diffraction pattern is modified (Fig. 1e, Supplementary Fig. 2). In this case, the modulator was made using 0.3 micron $Al_2O_3$ micropolish powder dispersed on kapton tape (see Methods for further details). Note that the modulator invariably contains more than 1 layer of 300 nm micropolish due to its fabrication. Prior to the measurement, we located a modulator region that induced significant changes to the exit beam, as shown in Fig. S2. The modulator features must be small enough relative to the exit wave's spatial size at the modulator plane (Supplementary Fig. 1) such that different portions of the exit wave are modulated differently. This depends on the crystal size, distance to the modulator plane, and x-ray wavelength. In practice, modulators of different sizes of $Al_2O_3$ micropolish can be prepared and checked to induce significant changes in the diffraction pattern (as in Fig. 1e). Using the algorithm detailed in the next section, the amplitude and phase of the modulator are reconstructed (Fig. 1f) using the known crystal that was reconstructed from the unmodulated data (Fig. 1c). Note that for this strategy to work, the modulated and unmodulated data must be measured at the same $\delta q_{z,i}$, which is easy to implement in practice. Also note that the modulator need not be perpendicular to the diffracted x-ray beam. Any deviation from a normal incidence angle will cause the modulator reconstruction to represent a projection of the modulator. Note that there are several experimental uncertainties that affect the modulator reconstruction, most notably the distance between the modulator and the sample. However, an effective representation of the modulator, meaning an object that captures the impact of the modulator on the exit wave, is really the key element. This effective representation allows the reconstruction of a single view of an extended object as discussed in the second half of the text.

(a) Initial preparation

1. Compute $\int \exp(i\delta q_{z,i} z)\rho(x,y,z)dz$ for all $\delta q_{z,i}$ for i=-60:1:60 and store

2. Fresnel propagate all of the 2D complex images to the modulator

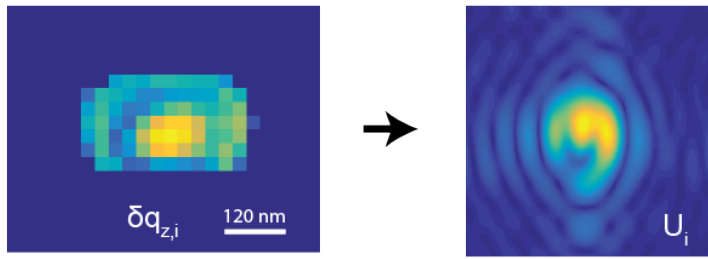

(b) Iterative loop to solve for the modulator

II. Fourier transform to propagate to the far field

I. For a given $\delta q_{z,i}$, apply multiplicative modulation

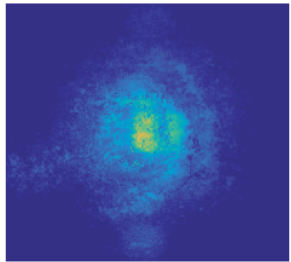 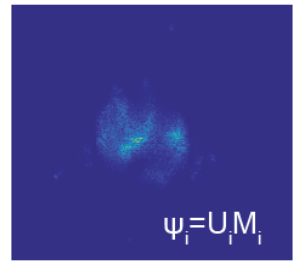

$\psi_i = U_i M_i$

Move to $\delta q_{z,i+1}$

III. Replace amplitudes by measured amplitudes

IV. Inverse Fourier transform and update modulator

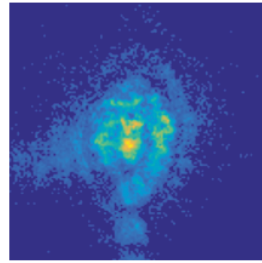

$$M_{i+1} = M_i + \alpha U_i^* \frac{\psi_i' - \psi_i}{\max(|U_i|^2)}$$

(c) Spatial schematic of the iterative steps

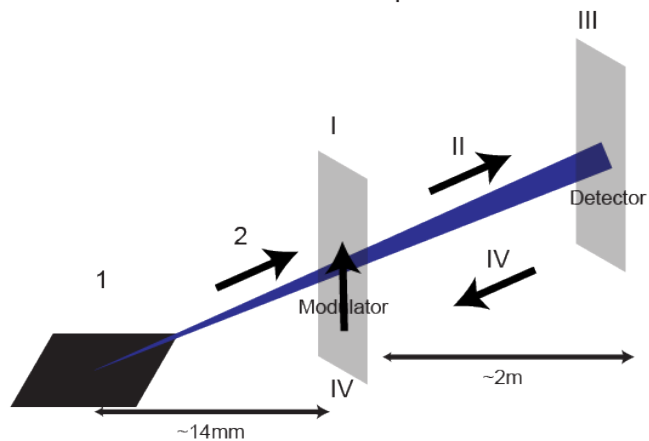

**Figure 2.** Iterative phase retrieval of the modulator structure. **(a)** The initial preparation before the iterative loop consists of 1) computing the complex 2D crystal projections for all $\delta q_{z,i}$ and 2) propagating these crystal projections to the modulator plane to form $U_i$. This is done for i=-60, -59, …, 60 in this case. **(b)** The iterative loop to solve for the modulator. In step I, a particular $U_i$ (connected to a particular $\delta q_{z,i}$) is chosen. The modulator interaction is modeled as $U_iM_i$ where $M_i$ is the current best guess for the modulator (a random initial guess is used). The modulator interaction causes both phase and amplitude changes. In step II, $\psi_i$ is propagated to the detector using a Fourier transform. In step III, the modulus constraint is applied: the measured amplitudes, $\sqrt{I_i}$, replace the current best guess for the amplitudes. In step IV, the inverse Fourier transform is used to propagate back to the modulator plane yielding $\psi'_i$. A ptychographic iterative engine-like update is used to update the best guess for the modulator. After step IV, the process is repeated for the next $\delta q_{z,i}$. **(c)** A spatial schematic of where each algorithmic step takes place along with the distances between the various planes for this particular experiment.

The key insight of the algorithm is that the diversity of the exit waves constrains an unknown modulator, similar to the concept of illumination diversity in ptychography. Figure 2 shows schematically how the algorithm works (MATLAB code is provided in the Supplementary Material). Note that all images are complex and only the intensity is shown. The first step is to take diffraction data for a particular crystal at different values of $\delta q_{z,i}$ without the modulator, and use this data to reconstruct the 3D complex crystal image by using standard phase retrieval algorithms (Fig. 1a-c). Consequently, the reference crystal exit wave can be computed from the 3D reconstructed image for the set of all $\delta q_{z,i}$ (Fig. 2a, Step 1). All of the 2D images are propagated using the Fresnel propagator to the modulator plane (Fig. 2a, Step 2). In our experiment, the modulator was approximately 14 mm away. While this has the advantage of being compatible with in situ experiments, the array size required to avoid aliasing while also sampling at the Nyquist frequency[36] for these experimental parameters is $10^4$x$10^4$. To decrease the array size, we used a larger real space sample pixel size of 28.7 nm. Further discussion regarding the array size and sampling is given in the Methods.

After Fresnel propagation to the modulator yields a set of $U_i$, a particular $U_i$ is chosen and the iterative loop to determine the modulator begins. The $U_i$ wavefield is multiplied by the complex modulator image to form $\psi_i$ (Fig. 2b Step I). $\psi_i$ is Fourier transformed to propagate the wavefield to the detector (~2m away) (Fig. 2b Step II). The amplitudes of the complex wavefield are replaced by the measured amplitudes ($\sqrt{I_i}$) of the diffraction data taken with the modulator in place (Fig. 2b Step III). This is known as the modulus projection step. The modulated data must be taken at the same set of $\delta q_{z,i}$ as the unmodulated data. An inverse Fourier transform is used to back propagate to the modulator plane to give $\psi'_i$ (Fig. 2b Step IV). The modulator is then updated using a ptychographic iterative engine-like update in which $U_i$ takes the role of the probe[31,37]:

$$M_{i+1} = M_i + \alpha U_i^* \frac{\psi'_i - \psi_i}{\max(|U_i|^2)} \quad (2)$$

In this expression, $M_i$ is the previous best guess for the modulator, $\alpha$ is the gradient step size that is set to 0.5, $U_i^*$ is the complex conjugate of $U_i$ (defined in Eq. (1)), $\psi_i = U_i M_i$, and $\psi'_i$ is the wavefield at the modulator plane after the modulus projection:

$$\psi'_i = \mathcal{F}o^{-1}(\sqrt{I_i} \cdot \frac{\mathcal{F}o(\psi_i)}{|\mathcal{F}o(\psi_i)|}) \quad (3)$$

where $I_i$ is the 2D modulated diffraction pattern at a particular $\delta q_{z,i}$ measured with the modulator in the exit beam, and $\mathcal{F}o$ is the 2D Fourier transform.

After the 2D complex image of the modulator for the given $\delta q_{z,i}$ is updated, the process is repeated for $\delta q_{z,i+1}$ (Fig. 2b, Step I). Fig. 2c shows schematically where the algorithm's iterative steps take place. Note that solving for the modulator requires iterating between the modulator plane and the detector plane only. No Fresnel propagation to the sample plane is required during the iterative loop. Using this algorithm, we solved for the 2D complex image of an unknown modulator, which is shown in Fig. 1f. Note that the $s_z$ projection image has a maximum amplitude ~10 due to the summing of multiple 2D slices. Thus, the maximum modulator amplitude of 2.5 should be considered relative to this number and can be normalized after the fact. Thus, we have demonstrated experimentally that an effective representation of an unknown modulator can be reconstructed using a known crystal. We now assume that the modulator is known and demonstrate with simulations how to image a targeted volume of material in an extended crystal using a single diffraction pattern.

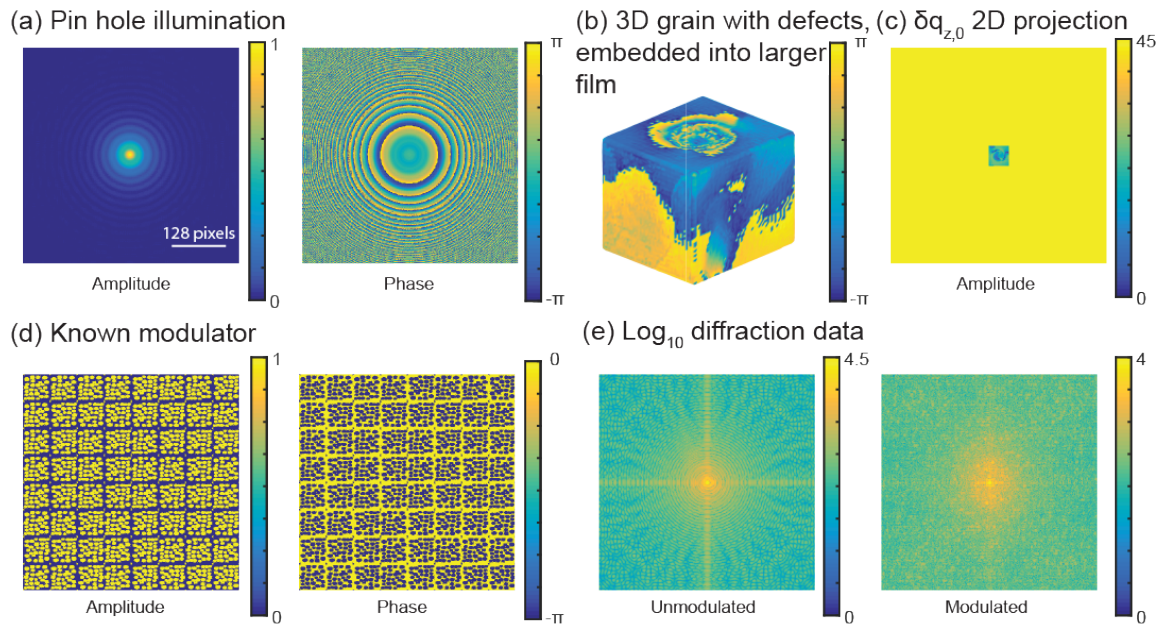

(a) Pin hole illumination
(b) 3D grain with defects, embedded into larger film
(c) $\delta q_{z,0}$ 2D projection
(d) Known modulator
(e) $Log_{10}$ diffraction data

**Figure 3.** A Bragg Coherent Modulation Imaging simulation. **(a)** The incident probe is computed from a coherently illuminated pinhole 1 mm from the sample plane using Fresnel propagation. The view shown is along the propagation axis. The scalebar applies to all images except the isosurface view in **(b)**. **(b)** Isosurface rendering of the 3D crystal with dislocations that is subsequently embedded into a larger crystal. The dislocation arrangement is computed from nanoindenting simulations. The extended sample is shown in **(c),** which shows the $\delta q_{z,0}$ projection. The sample is much larger than the beam. **(d)** The modulator considered in this case, which consists of unit amplitude, $-\pi$ phase shift regions. **(e)** The $\log_{10}$ of the diffraction amplitudes in the unmodulated and modulated case. As expected, the modulator increases the scattering diversity.

In BCMI, constraints during iterative phase retrieval are applied on 3 parallel planes: the sample, modulator, and detector plane. The probe in Eq. 1 is now modeled as the wavefield produced 1 mm downstream from a coherently illuminated, circular pinhole (Fig. 3a). The view shown is along the propagation axis. The sample considered is a nickel crystal that was subjected to nanoindentation using the atomistic simulations discussed previously[38,39] (Fig. 3b). The isosurface is drawn using the amplitude and the colormap projected onto the isosurface corresponds to the phase, which is proportional to the displacement of the atoms from their equilibrium positions. Dislocations nucleated in response to the nanoidentation can be identified as phase vortices (spiral regions of –π to π variation that cannot be removed by a global phase offset)[13,40–42]. The case we consider is this crystal embedded in an otherwise defect-free extended crystalline sample (Fig. 3c). Although the boundary between the defective and perfect film regions is discontinuous in phase, this numerical sample illustrates our concept.

Figure 3c shows the 2D projection of the crystal corresponding to $\delta q_{z,0}$, which is at the Bragg peak maximum. The modulator considered consists of a random tiling of unit amplitude, $-\pi$ phase shift structures (Fig. 3d). There are several methods available for constructing this type of modulator, including depositing films on TEM grids, optical lithography, and electroformed meshes. When considering cost, electroformed meshes appear to be the most promising direction. Finally, the diffraction data for the unmodulated and modulated exit wave fields are shown (Fig. 3e). In the following simulations, we assume that the probe and modulator are known.

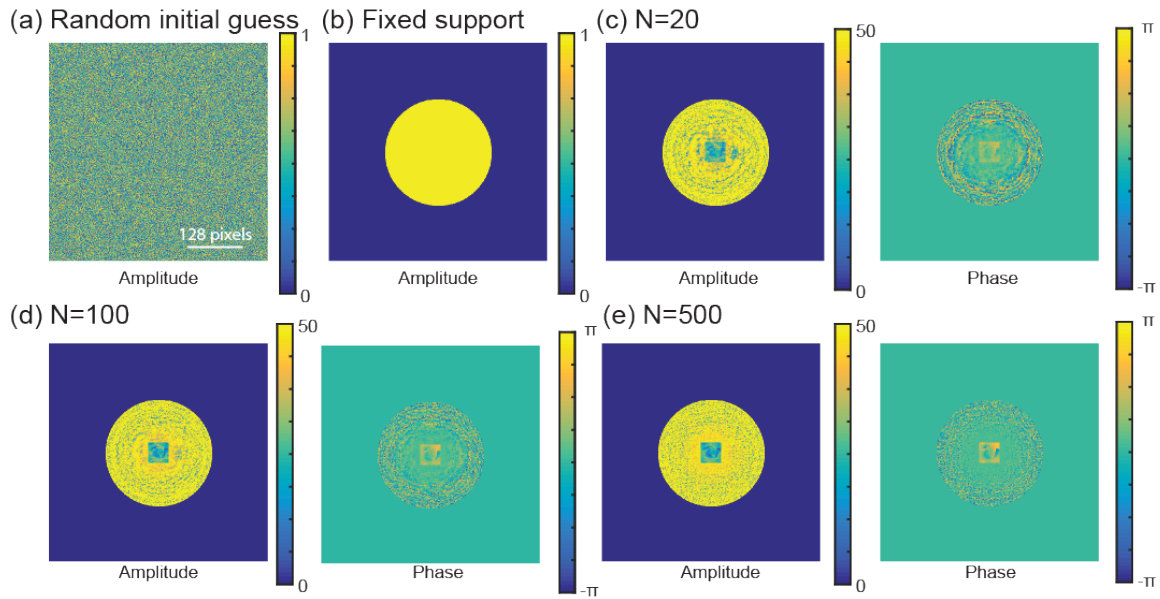

**Figure 4.** Bragg Coherent Modulation Imaging reconstructions: a single view from a single diffraction pattern/single position of an extended object. **(a)** A random array for both the amplitude (shown) and phase (not shown) of the sample exit wave is used as the initial guess. The scalebar applies to all images. **(b)** A fixed, real support is used as the real space constraint. The support is approximately twice the probe size. **(c)** The best guess for the object after 20 iterations. The structure in the amplitude and phase is beginning to emerge at the array center. **(d)** The best guess for the object after 100

iterations. **(e)** The best guess for the object after 500 iterations. In the image center, both the amplitude and the phase have converged to their true values. Note that the object is indeed extended, with non-zero amplitude throughout the entire computational array, simulating a continuous thin film with a defective region in the center of the field of view.

In our algorithm, only a single diffraction pattern is used to recover a single 2D projection of the crystal in a field of view corresponding to the probe size. We have chosen the $\delta q_{z,0}$ projection. $U = U_0$ is defined in Eq. (1) with $P$ being the wavefield from the pinhole. A random initial guess for both the sample exit wave amplitude and phase is used (Fig. 4a). This guess is propagated to the modulator plane where it interacts with the modulator via

$$\psi = UM \quad (4)$$

$\psi$ is propagated to the detector via the Fourier transform. The modulus constraint is used, where the amplitude of the wave is replaced by the square root of the measured data and then back propagated to the modulator plane:

$$\psi' = \mathcal{F}o^{-1}(\sqrt{I} \cdot \frac{\mathcal{F}o(\psi)}{|\mathcal{F}o(\psi)|}) \quad (5)$$

The modulator is removed from the wave through division:

$$U = \frac{\psi'}{M} \quad (6)$$

This wave is back propagated to the sample plane where a finite support constraint is applied to the sample exit wave (Fig. 4b). The support is set to be twice as large as the probe and is not updated during the iterative loop. This is to demonstrate that a very loose support does not have detrimental effects. The process is repeated until both the finite support and modulus constraint have been applied 1000 times. The convergence process is shown in Figs. 4c-e. After 500 iterations, the amplitude and phase are near their true values. The MATLAB code used is available in the Supplemental Material. The final reconstruction after 1000 iterations is shown next to the true solution in Figure 5. Note that while the object may appear to be finite, there is non-zero amplitude throughout the entire computational array. Supplementary Fig. 3 shows the results from considering additional dislocations tiling the computational array. Supplementary Fig. 4 shows the relative sizes of the support, object, probe, and reconstruction. The reconstruction quality degrades rapidly outside of significant probe intensity.

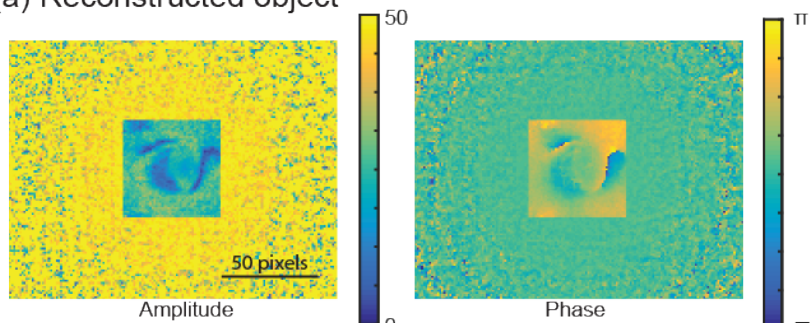

(a) Reconstructed object

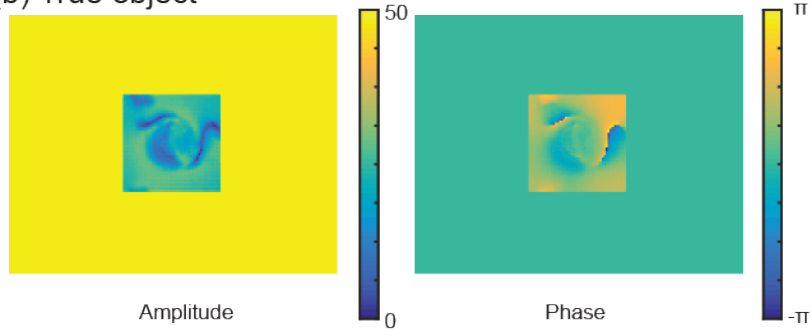

(b) True object

**Figure 5.** Comparison of the reconstructed image with the true image**. (a)** The reconstructed image after 1000 iterations of the algorithm described in the text. The scalebar applies to all images. **(b)** The true object. The agreement is very good in the center of the array and becomes worse at the edges. A region of an extended sample has successfully been imaged using a single diffraction pattern.

Figure 5 shows a comparison of the final reconstructed object with the true solution. There is quantitative agreement at the reconstruction center. However, the reconstruction quality degrades further away from the center. To quantitatively assess the agreement, we compute the average over a 3-pixel annulus at a given radius of $|\angle \delta q_{\{z,0,\text{guess}\}} - \angle \delta q_{\{z,0,\text{true}\}}|$ to assess agreement in the phase and $||\delta q_{\{z,0,\text{guess}\}}| - |\delta q_{\{z,0,\text{true}\}}||$ to assess agreement in the amplitude. A plot of these quantities as a function of radius (zero defined at the center of the array) for the object considered in Figs. 3-5 and the object considered in Supplementary Fig. 3 is shown in Supplementary Fig. 5.

The degrading reconstruction quality at larger radii is related to the spatial extent of the probe. However, because only a single diffraction pattern at a single position is required, BCMI has significantly improved temporal resolution compared to Bragg ptychography for features that fit within the beam footprint. Should an extended view be required, the sample can be scanned in the beam and a reconstruction at each position done independently. Additionally, the 3D reconstruction can be assembled from 2D projections, each of which uses the same probe-sample intersection volume[34]. We also show the results of a single view imaging simulation of an extended sample with many defects in Supplementary Fig. 3 to demonstrate that BCMI works in the presence of many dislocations. We use this case to demonstrate that the reconstruction with no modulator and with an amplitude-only modulator fails to converge to the object (Supplementary Fig. 6). This highlights the importance of the modulator and in particular the phase shift it induces in the exit wave.

We also investigated whether shrink-wrapping[46,47] the support will improve the reconstruction. We used a Gaussian shrink-wrap function with a threshold of 0.1 and a standard deviation of 1. For both types of objects investigated (a single defect structure embedded in an otherwise perfect crystal as in Figs. 3-5 or many defects embedded in a crystal as in Supplementary Figs. 3, 6), shrink-wrapping degrades the reconstruction accuracy (Supplementary Fig. 7). BCMI shares some similarities with the single-view and ptychographic Fresnel CDI[48,49]. In all techniques, accurate propagation of the wavefront using the Fresnel propagator is essential. However, BCMI does not require exact knowledge of the support size or sample translation in the x-ray beam to converge to the correct reconstruction. Finally, like all coherent-based imaging methods that rely on phase retrieval, we expect there is a limit to the number of defects and/or strain amount that this method can accurately reconstruct.

We have developed Bragg coherent modulation imaging (BCMI). We demonstrated how to reconstruct an unknown modulator using a known crystal and how a known modulator can be used for single-view imaging of an extended sample. The single-view imaging experiment will use a modulator fixed to the detector arm so that the exit waves from the isolated crystal and the extended sample travel through the same portion of the modulator. The modulator can be fabricated by sputtering micron thick layers of gold on holey silicon nitride membranes or through lithographic fabrication. These results pave the way for fast imaging of defect dynamics in extended samples in reactive environments, which will be a crucial tool to understanding defect-function and defect-stability relationships at the nanoscale.

**Methods:**

**Gold Nanocrystal Synthesis**
Gold nanocrystals were synthesized using a dewetting technique[50]. Briefly, 20 nm of Au was deposited onto a Si wafer that had a $SiO_2$ layer grown on it by heating in air. The as-deposited sample was heated to 950° C in a tube furnace for 2 hours. This caused the film to dewet and form isolated gold nanoparticles.

**BCDI experiments**

Experiments were performed at Sector 34-ID-C of the Advanced Photon Source at Argonne National Laboratory. A double crystal monochromator was used to select E=8.919 keV x-rays with 1 eV bandwidth and longitudinal coherence length of 0.7 $\mu m$. A set of Kirkpatrick-Baez mirrors was used to focus the beam to $0.6 \times 0.8\ \mu m^2$ (vertical x horizontal). The rocking curve around the Au (111) Bragg peak was collected by recording 2D coherent diffraction patterns with an x-ray sensitive area detector (Medipix2/Timepix, 512x512 pixels, each pixel 55µm x 55µm). It was placed a distance of 2.2 m away from the sample and an evacuated flight tube was inserted between the modulator and the detector. A total of 121 patterns were collected for a single 3D rocking scan over a total angular range of ($\Delta\theta = \pm 0.3°$). Each 3D data set took approximately 10 minutes.

**Standard Phase retrieval**

The phase retrieval code is adapted from published work[51,52]. The hybrid input-output[20,53] and error reduction algorithms were used for all reconstructions. A total of 1050 iterations, consisting of alternating 40 iterations of the hybrid input-output algorithm with 10 iterations of the error reduction algorithm, were run for 10 reconstructions beginning from random phases. The best reconstruction, quantified by the smallest sharpness metric, was then used in conjunction with another random phase start as a seed for another 10 random starts. The sharpness metric is the sum of the absolute value of the reconstruction raised to the 4$^{th}$ power. 10 generations were used in this guided algorithm[54].

**Modulator construction**

The modulator was constructed by dispersing $Al_2O_3$ 0.3 micron micropolish between two pieces of kapton tape. This was placed downstream of the sample by mounting the modulator on the optical microscope used for sample alignment.

**Modulated phase retrieval: reconstruction of the modulator**

Please see the Solve_modulator_knowncrystal_main.m file and the associated .mat files provided.

## Array size required for alias-free propagation

The array size for required to avoid aliasing while also sampling at the Nyquist frequency is[36]

$$N = \frac{z_{mod}\lambda}{\Delta x_{sample}\Delta x_{modulator}}$$

where $z_{mod}$ is the sample to modulator distance, $\lambda$ is the x-ray wavelength, $\Delta x_{sample}$ is the pixel size in the sample image, and $\Delta x_{modulator}$ is the pixel size in the modulator image. Here we assume equal pixel sizes in the sample and modulator image. This image pixel size assuming a square array is set by experimental parameters:

$$\Delta x_{sample} = \frac{z_{detector}\lambda}{D_{detector}}$$

where $z_{detector}$ is the distance from the sample to the detector, $\lambda$ is the x-ray wavelength, and $D_{detector}$ is the size of the detector (the number of pixels in one dimension times the pixel size, for example 256 pixels times 55 micron pixels). For a typical BCDI experiment, $\Delta x_{sample} \sim 10\ nm$, $\lambda \sim 1\ \text{Å}$, and thus for $z_{mod} \sim 10\ mm$ the array size is $\sim 10^4$ in both dimensions. For this work, we used a larger real space sample pixel size (28.7 nm) to reduce the array size to 1794x1794 to reduce memory requirements during the modulator reconstruction.

## Modulated phase retrieval: reconstruction of the object

Please see the singleviewBraggptycho_object_reconstruction_main.m file and the associated .mat files provided.

**Data availability statement:** The data reported in this paper are included in the supplementary material. All reconstruction code is included in the supplementary material.

## References:

1. Abbey, B. From Grain Boundaries to Single Defects: A Review of Coherent Methods for Materials Imaging in the X-ray Sciences. *JOM* **65,** 1183–1201 (2013).

2. Thibault, P., Elser, V., Jacobsen, C., Shapiro, D. & Sayre, D. Reconstruction of a yeast cell from X-ray diffraction data. *Acta Crystallogr. A.* **62,** 248–61 (2006).

3. Chapman, H. N. *et al.* Femtosecond diffractive imaging with a soft-X-ray free-electron laser. *Nat. Phys.* **2,** 839–843 (2006).

4. Pfeifer, M. A., Williams, G. J., Vartanyants, I. A., Harder, R. & Robinson, I. K. Three-dimensional mapping of a deformation field inside a nanocrystal. *Nature* **442,** 63–6 (2006).

5. Robinson, I. & Harder, R. Coherent X-ray diffraction imaging of strain at the nanoscale. *Nat. Mater.* **8,** 291–8 (2009).

6. Favre-Nicolin, V. *et al.* Analysis of strain and stacking faults in single nanowires using Bragg coherent diffraction imaging. *New J. Phys.* **12,** (2010).

7. Haag, S. T. *et al.* Anomalous coherent diffraction of core-shell nano-objects: A methodology for determination of composition and strain fields. *Phys. Rev. B - Condens. Matter Mater. Phys.* **87,** 1–16 (2013).

8. Hruszkewycz, S. O. *et al.* High-resolution three-dimensional structural microscopy by single-angle Bragg ptychography. *Nat. Mater.* **16,** 1–10 (2016).

9. Holt, M. V. *et al.* Strain imaging of nanoscale semiconductor heterostructures with x-ray bragg projection ptychography. *Phys. Rev. Lett.* **112,** 1–6 (2014).

10. Godard, P. *et al.* Three-dimensional high-resolution quantitative microscopy of extended crystals. *Nat. Commun.* **2,** 568 (2011).

11. Yau, A., Cha, W., Kanan, M. W., Stephenson, G. B. & Ulvestad, A. Bragg Coherent Diffractive Imaging of Single-Grain Defect Dynamics in Polycrystalline Films. *Science (80-. ).* **356,** 739–742 (2017).

12. Ulvestad, A. *et al.* Topological defect dynamics in operando battery nanoparticles. *Science* **348,** 1344–1347 (2015).

13. Ulvestad, A. *et al.* Three-dimensional imaging of dislocation dynamics during the hydriding phase transformation. *Nat. Mater.* **16,** 565–571 (2017).

14. Xiong, G., Clark, J. N., Nicklin, C., Rawle, J. & Robinson, I. K. Atomic Diffusion within Individual Gold Nanocrystal. *Sci. Rep.* **4,** 6765 (2014).

15. Clark, J. N. *et al.* Three-dimensional imaging of dislocation propagation during crystal growth and dissolution. *Nat. Mater.* **14,** 780–784 (2015).

16. Lawrence, N. J. *et al.* Defect engineering in cubic cerium oxide nanostructures for catalytic oxidation. *Nano Lett.* **11,** 2666–2671 (2011).

17. Seebauer, E. G. & Noh, K. W. Trends in semiconductor defect engineering at the nanoscale. *Mater. Sci. Eng. R Reports* **70,** 151–168 (2010).

18. Wu, J. *et al.* Surface lattice-engineered bimetallic nanoparticles and their catalytic properties. *Chem. Soc. Rev.* **41,** 8066–74 (2012).

19. Ben, X. & Park, H. S. Strain engineering enhancement of surface plasmon polariton propagation lengths for gold nanowires. *Appl. Phys. Lett.* **102,** 41909 (2013).

20. Fienup, J. R. Phase retrieval algorithms: a comparison. *Appl. Opt.* **21,** 2758–69

(1982).

21. Adams, D. E. *et al.* A generalization for optimized phase retrieval algorithms. *Opt. Express* **20,** 24778 (2012).

22. Marchesini, S. Phase retrieval and saddle-point optimization. *J. Opt. Soc. Am. A. Opt. Image Sci. Vis.* **24,** 3289–3296 (2007).

23. Rodenburg, J. M. Ptychography and related diffractive imaging methods. *Adv. Imaging Electron Phys.* **150,** 87–184 (2008).

24. Maiden, A. M., Humphry, M. J., Zhang, F. & Rodenburg, J. M. Superresolution imaging via ptychography. *J. Opt. Soc. Am. A. Opt. Image Sci. Vis.* **28,** 604–612 (2011).

25. Marchesini, S. A unified evaluation of iterative projection algorithms for phase retrieval. *Rev. Sci. Instrum.* **78,** 11301 (2007).

26. Miao, J. W., Charalambous, P., Kirz, J. & Sayre, D. Extending the methodology of X-ray crystallography to allow imaging of micrometre-sized non-crystalline specimens. *Nature* **400,** 342–344 (1999).

27. Peterson, I., Harder, R. & Robinson, I. K. Probe-diverse ptychography. *Ultramicroscopy* **171,** 77–81 (2016).

28. Beck, A. & Eldar, Y. C. Sparsity constrained nonlinear optimization: optimality conditions and algorithms ∗. *SIAM J. Optim.* **23,** 1480–1509 (2013).

29. Tripathi, A., McNulty, I., Munson, T. & Wild, S. M. Single-view phase retrieval of an extended sample by exploiting edge detection and sparsity. *Opt. Express* **24,** 24719–24738 (2016).

30. Zhang, F. *et al.* Phase retrieval by coherent modulation imaging. *Nat. Commun.* **7,** 13367 (2016).

31. Zhang, F. & Rodenburg, J. M. Phase retrieval based on wave-front relay and modulation. *Phys. Rev. B - Condens. Matter Mater. Phys.* **82,** 1–4 (2010).

32. Ulvestad, A. *et al.* Coherent diffractive imaging of time-evolving samples with improved temporal resolution. *Phys. Rev. B* **93,** 184105 (2016).

33. Vartanyants, I. & Robinson, I. Partial coherence effects on the imaging of small crystals using coherent x-ray diffraction. *J. Phys. Condens. …* **10593,** (2001).

34. Cha, W. *et al.* Three Dimensional Variable-Wavelength X-Ray Bragg Coherent Diffraction Imaging. *Phys. Rev. Lett.* **117,** 1–5 (2016).

35. Vartanyants, I. A. & Oleksandr, Y. in *X-Ray Diffraction: Modern Experimental Techniques* (eds. Seeck, O. H. & Murphy, B. M.) 1–51 (Pan Stanford, 2015).

36. Mas, D., Garcia, J., Ferreira, C., Bernardo, L. M. & Marinho, F. Fast algorithms for free-space diffraction patterns calculation. *Opt. Commun.* **164,** 233–245 (1999).

37. Maiden, A. M., Johnson, D. & Li, P. Further improvements to the ptychographical iterative engine. *Optica* **4,** 736–745 (2017).

38. Dupraz, M., Beutier, G., Rodney, D., Mordehai, D. & Verdier, M. Signature of dislocations and stacking faults of face-centred cubic nanocrystals in coherent X-ray diffraction patterns: a numerical study. *J. Appl. Crystallogr.* **48,** 621–644

(2015).

39. Chang, H.-J., Fivel, M., Rodney, D. & Verdier, M. Multiscale modelling of indentation in FCC metals: From atomic to continuum. *Comptes Rendus Phys.* **11,** 285–292 (2010).

40. Jacques, V. L. R. *et al.* A coherent way to image dislocations. 8 (2010). at <http://arxiv.org/abs/1008.5250>

41. Jacques, V. L. R. *et al.* Bulk Dislocation Core Dissociation Probed by Coherent X Rays in Silicon. *Phys. Rev. Lett.* **106,** 65502 (2011).

42. Takahashi, Y. *et al.* Bragg x-ray ptychography of a silicon crystal: Visualization of the dislocation strain field and the production of a vortex beam. *Phys. Rev. B* **87,** 121201 (2013).

43. Als-nielsen, J. & Mcmorrow, D. *Elements of modern X-ray physics*. (John Wiley & Sons, Inc., 2011).

44. Irvine, S. C. *et al.* Assessment of the use of a diffuser in propagation-based x-ray phase contrast imaging. **18,** 13478–13491 (2010).

45. Awaji, M., Suzuki, Y., Takeuchi, A. & Takano, H. Zernike-type X-ray imaging microscopy at 25 keV with Fresnel zone plate optics research papers. 125–127 (1998).

46. Chapman, H., Barty, A. & Marchesini, S. High-resolution ab initio three-dimensional x-ray diffraction microscopy. *JOSA A* **23,** (2006).

47. Marchesini, S., He, H. & Chapman, H. X-ray image reconstruction from a diffraction pattern alone. *Phys. Rev. B* **68,** 140101 (2003).

48. Williams, G. J. *et al.* Fresnel Coherent Diffractive Imaging. *Phys. Rev. Lett.* **97,** 1–4 (2006).

49. Vine, D. J. *et al.* Ptychographic Fresnel coherent diffractive imaging. *Phys. Rev. A* **80,** 1–5 (2009).

50. Kracker, M., Wisniewski, W. & Rüssel, C. Textures of Au, Pt and Pd/PdO nanoparticles thermally dewetted from thin metal layers on fused silica. *RSC Adv.* **4,** 48135–48143 (2014).

51. Yang, W. *et al.* Coherent diffraction imaging of nanoscale strain evolution in a single crystal under high pressure. *Nat. Commun.* **4,** 1680 (2013).

52. Clark, J. N., Huang, X., Harder, R. & Robinson, I. K. High-resolution three-dimensional partially coherent diffraction imaging. *Nat. Commun.* **3,** 993 (2012).

53. Fienup, J. R., Wackerman, C. C. & Arbor, A. Phase-retrieval stagnation problems and solutions. *J. Opt. Soc. Am. A* **3,** 1897–1907 (1986).

54. Chen, C.-C., Miao, J., Wang, C. & Lee, T. Application of optimization technique to noncrystalline x-ray diffraction microscopy: Guided hybrid input-output method. *Phys. Rev. B* **76,** 64113 (2007).

**Acknowledgements:**

Work was supported by the U.S. DOE, Basic Energy Sciences, Materials Sciences and Engineering Division (Phase retrieval algorithm development and testing). The synchrotron x-ray imaging experiment used resources of the Advanced Photon Source, funded by the U.S. Department of Energy (DOE) Office of Science User Facilities (OSUF) operated for the DOE Office of Science by Argonne National Laboratory under Contract No. DE-AC02-06CH11357.

**Author contributions:** A.U. designed the phase retrieval algorithm, modulator experiment, performed the simulations, and wrote the manuscript. W.C. built the modulator. All authors participated in the experiment and edited the manuscript.

**Additional Information**

**Supplementary information** is available in the online version of the paper. Reprints and permissions information is available online at www.nature.com/reprints. Correspondence and requests for materials should be addressed to A.U

**Competing Financial Interests:** The authors declare no competing financial interests.